\documentclass[onecolumn, floatfix, preprintnumbers, eqsecnum, letterpaper, superscriptaddress, nofootinbib]{revtex4}
\usepackage{graphicx}
\usepackage{microtype}
\usepackage{amsmath}
\usepackage{amssymb}
\usepackage{subfigure}
\usepackage{hyperref}
\usepackage{url}
\usepackage{xcolor}
\usepackage{color}
\usepackage{mathrsfs}
\usepackage{calrsfs}
\usepackage{amsfonts}
\usepackage{eufrak}
\usepackage{tabularx}
\usepackage{eucal}
\usepackage{latexsym}
\usepackage{ragged2e}
\usepackage{epsfig}
\usepackage{textcomp}
\usepackage{caption}
\usepackage{subfigure}
\usepackage{marvosym}
\usepackage{ifsym}
\usepackage{lipsum}
\usepackage{appendix}
\usepackage{bm}
\usepackage{mathrsfs}

\DeclareCaptionJustification{justified}{\leftskip=0pt \rightskip=0pt \parfillskip=0pt plus 1fil}
\definecolor{vividviolet}{rgb}{0.62, 0.0, 1.0}
\definecolor{amaranth}{rgb}{0.9, 0.17, 0.31}
\definecolor{palatinateblue}{rgb}{0.15, 0.23, 0.89}
\definecolor{brightpink}{rgb}{1.0, 0.0, 0.5}
\definecolor{cornflowerblue}{rgb}{0.39, 0.58, 0.93}
\definecolor{deepcarminepink}{rgb}{0.94, 0.19, 0.22}
\definecolor{radicalred}{rgb}{1.0, 0.21, 0.37}

\hypersetup{ linktoc=all,
    colorlinks, linkcolor={palatinateblue},
    citecolor={brightpink}, urlcolor={amaranth}
}

\graphicspath{{Images/}}

\graphicspath{{Images/}}

\def\@fnsymbol#1{\ensuremath{\ifcase#1\or \ddagger \or  $\textleaf$  \or \dagger
\else\@ctrerr\fi}}%

\makeatother

\def\sideremark#1{\ifvmode\leavevmode\fi\vadjust{\vbox to0pt{\vss
 \hbox to 0pt{\hskip\hsize\hskip1em
 \vbox{\hsize1.3cm\tiny\raggedright\pretolerance10000
 \noindent #1\hfill}\hss}\vbox to8pt{\vfil}\vss}}}%

\def\beq{\begin{equation}}
\def\eeq{\end{equation}}

\newcommand{\od}{\mathrm{d}}

\begin{document}

\title{The $P$-$V$ Phase Transition of the FRW Universe}

\author{Shi-Bei Kong}
\email{kongshibei@nuaa.edu.cn}
\affiliation{College of Physics, Nanjing University of Aeronautics and Astronautics, Nanjing, 211106, China}

\author{Haximjan Abdusattar}
\email{axim@nuaa.edu.cn}
\affiliation{College of Physics, Nanjing University of Aeronautics and Astronautics, Nanjing, 211106, China}

\author{Yihao Yin}
\email{yinyihao@nuaa.edu.cn}
\affiliation{College of Physics, Nanjing University of Aeronautics and Astronautics, Nanjing, 211106, China}

\author{Hongsheng Zhang}
\email{sps$_$zhanghs@ujn.edu.cn}
\affiliation{School of Physics and Technology, University of Jinan, 336 West Road of Nan Xinzhuang, Jinan, Shandong 250022, China}
\affiliation{Key Laboratory of Theoretical Physics, Institute of Theoretical Physics, Chinese Academy of Sciences, Beijing 100190, China}

\author{Ya-Peng Hu $^{\text{\Letter}}$}
\email{huyp@nuaa.edu.cn}
\affiliation{College of Physics, Nanjing University of Aeronautics and Astronautics, Nanjing, 211106, China}
\affiliation{Key Laboratory of Aerospace Information Materials and Physics (NUAA), MIIT, Nanjing 211106, China}

{\let\thefootnote\relax\footnotetext{\vspace*{0.2cm}$^{\text{\Letter}}$ Corresponding Author}}

\begin{abstract}

We define thermodynamic pressure $P$ by work density $W$ as the conjugate quantity of thermodynamic volume $V$ from field equation.  We derive the equations of state $P$=$P(V, T)$ for the Friedmann-Robertson-Walker (FRW) universe in Einstein gravity and a modified gravity respectively. We find that the equation of state from Einstein gravity
shows no $P$-$V$ phase transition, while the equation of state from the modified gravity does, where the critical exponents are the same as those in mean field theory.

\end{abstract}

\maketitle

\section{Introduction}

The Friedmann-Robertson-Walker (FRW) universe is a thermodynamic system similar to a black hole.
In theory, a black hole has many well-known properties, e.g.\ the existence of a horizon, Hawking temperature and Hawking radiation \cite{Hawking:1975vcx}, Bekenstein-Hawking entropy, quasi-local energy, unified first law \cite{Hayward:1993wb,Hayward:1997jp}, etc. These mentioned properties are all shared by the FRW universe \cite{Padmanabhan:2003gd,Cai:2005ra,Akbar:2006er,Akbar:2006kj,Cai:2006rs,Cai:2008gw,Zhu:2009wa,Hu:2010tx,Gong:2007md}.

On the other hand, some other properties of black holes have not yet been known to exist for the FRW universe. Take the asymptotically AdS black hole as an example, in both Einstein gravity and modified gravity, it usually has a thermodynamic equation of state $P=P(V,T)$, where $V$ is the thermodynamic volume, $T$ is the Hawking temperature, and $P$ is the thermodynamic pressure that is usually defined by the cosmological constant  \cite{Kastor:2009wy,Dolan:2011xt,Kubiznak:2012wp,Cai:2013qga,Kastor:2014dra,Carlip:2014pma}
\footnote{However, it should be noted that from holography or AdS/CFT, the cosmological constant $\Lambda$ is related with the rank $N$ of a gauge group (or number of colors) \cite{Kastor:2009wy}, so treating the cosmological ``constant" in the bulk as a thermodynamical ``variable" corresponds to treating the number of colors in the dual boundary theory as a thermodynamical variable \cite{Kastor:2009wy,Johnson:2014yja,Dolan:2014cja,Kastor:2014dra}, which changes the theory and does not belong to the standard thermodynamical method \cite{Karch:2015rpa}. }.
In some situations such an equation is characterized the $P$-$V$ phase transition \cite{Xu:2015rfa,Hendi:2015eca,Hendi:2017fxp,Nam:2018ltb,Hu:2018qsy,Hu:2020pmr}
with a critical point in the $P$-$V$ phase diagram.
However, for the FRW universe, so far there has been rarely any investigations on the equation of state, $P$-$V$ phase transitions and criticality.
In a recent paper \cite{Abdusattar:2021wfv}, the equation of state for the FRW universe with a perfect fluid in Einstein gravity has been constructed, but no $P$-$V$ phase transition was found.
\footnote{Note that investigations have shown that the existence of an apparent horizon is the main cause of having a self-consistent thermodynamics \cite{Hayward:1993wb,Padmanabhan:2003gd,Cai:2005ra,Gong:2007md}. Although the FRW thermodynamics was historically inspired by that of black holes, the self-consistency of the former in principle does not rely on the latter.}

Out of curiosity, in this paper we would like to find a reasonable way to construct an equation of state that can describe phase transitions of the FRW universe. The construction of such an equation of state depends on: (i) the definition of the thermodynamic quantities ($P$, $V$, $T$); (ii) the choice of the gravitational theory; and (iii) the properties of the matter field or source. The basic setup for this paper is the following:
\linebreak
For (i), we follow the definitions of the thermodynamic volume $V$ and the Hawking temperature $T$ in \cite{Akbar:2006kj,Cai:2006rs}, but very importantly we change the definition of $P$.
From the first law of thermodynamics for the FRW universe in Einstein gravity and many modified theories of gravity,
we find that the work density $W$ of the matter field is the conjugate variable of the thermodynamic volume,
so it should be defined as the thermodynamic pressure
\begin{alignat}{1}
P\equiv W=-\frac{1}{2}h_{ab}T^{ab},
\end{alignat}
where $h_{ab}$ and $T^{ab}$ are the $0,1$-components of the metric and the stress-tensor \cite{Akbar:2006kj} with $a,b=0,1, (x^0=t, x^1=r)$. For (ii), after a brief discussion on Einstein gravity, we will mainly focus on a modified theory of gravity
that belongs to the Horndeski class, because it gives interesting Friedmann's equations. For (iii) we treat the matter field as a perfect fluid as is usually done in standard cosmology.

This paper is organized as follows. In Sec.II, as a warmup exercise, we construct the equation of state for the FRW universe in Einstein gravity, which does not show $P$-$V$ phase transition.
In Sec.III, we obtain the equation of state for the FRW universe in a modified gravity that belongs to the Horndeski class.
In Sec.IV, we show that $P$-$V$ phase transition of the FRW universe exists in the modified gravity.
Sec.V is for conclusions and discussions.
In this paper, we use units $c=G=\hbar=k_B=1$.

\section{Einstein Gravity: No $P$-$V$ Phase Transition}

In this section, we first show that the thermodynamic pressure $P$ can be defined by the work density $W$ for the FRW universe in Einstein gravity, and then construct the equation of state $P=P(V,T)$, which shows no $P$-$V$ phase transition.

\bigskip

In the co-moving coordinate system $\{t,r,\theta,\varphi\}$, the line-element of the FRW universe can be written as
\begin{alignat}{1}
\od s^2=-\od t^2+a^2(t)\left[\frac{\od r^2}{1-kr^2}+r^2(\od\theta^2+\sin^2\theta\od\varphi^2)\right], \label{LE}
\end{alignat}
where $a(t)$ is the time-dependent scale factor
\footnote{It means that the FRW universe belongs to dynamical spacetimes, and has no global timelike Killing vector field,
which makes it difficult to define conserved charges using the method for stationary spacetimes.
In the Appendix, we show how to define conserved quantities in the FRW universe and why the FRW universe can be regarded as a quasi-equilibrium system.},
$k$ is the spatial curvature. The stress-tensor of a perfect fluid is usually written as
\begin{alignat}{1}
T_{\mu\nu}=(\rho_m+p_m)u_{\mu}u_{\nu}+p_m g_{\mu\nu},\label{ST}
\end{alignat}
where $\rho_m$ is the energy density, $p_m$ is the pressure and $u^{\mu}$ is the 4-velocity of the perfect fluid.
In Einstein gravity, the line-element (\ref{LE}) and the stress-tensor (\ref{ST}) satisfy Einstein field equations
\begin{alignat}{1}
R_{\mu\nu}-\frac{1}{2}g_{\mu\nu}R=8\pi T_{\mu\nu},\label{FE}
\end{alignat}
which gives the Friedmann's equations
\begin{alignat}{1}
H^2+\frac{k}{a^2}=\frac{8\pi}{3}\rho_m, \quad \dot{H}-\frac{k}{a^2}=-4\pi(\rho_m+p_m), \label{FE}
\end{alignat}
where $H:=\dot{a}(t)/a(t)$ is the Hubble parameter.

For convenience in the following discussion, we introduce another form of the line-element (\ref{LE})
\begin{alignat}{1}
\od s^2=h_{ab}\od x^a\od x^b+R^2(\od\theta^2+\sin^2\theta\od\varphi^2),
\end{alignat}
where $a,b=0,1, x^0=t, x^1=r$ and $R\equiv a(t)r$ is the physical radius
\footnote{The metric of the FRW universe in physical coordinates are also given in the Appendix.}
of the FRW universe. The apparent horizon of the FRW universe is defined by the following condition \cite{Cai:2005ra}
\begin{alignat}{1}
h^{ab}\partial_a R\partial_b R=0,
\end{alignat}
which can be easily solved
\begin{alignat}{1}
R_A=\frac{1}{\sqrt{H^2+\frac{k}{a^2}}}. \label{AH}
\end{alignat}
From the above expression, one can get a useful relation
\begin{alignat}{1}
\dot{R}_A=-HR^3_A\left(\dot{H}-\frac{k}{a^2}\right). \label{dot}
\end{alignat}

The surface gravity at the apparent horizon of the FRW universe is given as  \cite{Cai:2005ra}
\begin{alignat}{1}
\kappa=-\frac{1}{R_A}\left(1-\frac{\dot{R}_A}{2HR_A}\right), \label{sg}
\end{alignat}
and we treat $\dot{R}_A$ as a small quantity, so that the surface gravity $\kappa$ is negative,
i.e.\ the apparent horizon of the FRW universe is an inner trapping horizon \cite{Hayward:1993wb} .
This surface gravity has a simple relation with the Ricci scalar of the FRW universe
\begin{alignat}{1}
\kappa=-\frac{R_A R}{12}.
\end{alignat}
The Kodama-Hayward temperature is defined from the surface gravity (\ref{sg})
\begin{alignat}{1}
T\equiv\frac{|\kappa|}{2\pi}=\frac{1}{2\pi R_A}\left(1-\frac{\dot{R}_A}{2HR_A}\right). \label{KHT}
\end{alignat}
Furthermore, in Einstein gravity we have Bekenstein-Hawking entropy
\begin{alignat}{1}
S\equiv&\frac{A}{4}=\pi R_A^2,
\end{alignat}
and Misner-Sharp energy \cite{Maeda:2007uu,Cai:2009qf}
\begin{alignat}{1}
M\equiv\frac{R_A}{2}
\end{alignat}
for the FRW universe.

From (\ref{FE}), (\ref{AH}) and (\ref{dot}), one can express $\rho_m$ and $p_m$ in terms of $R_A$ and $\dot{R}_A$, $i.e.$
\begin{alignat}{1}
\rho_m=\frac{3}{8\pi R^2_A}, \quad p_m=\frac{\dot{R}_A}{4\pi HR^3_A}-\frac{3}{8\pi R^2_A}. \label{rhop}
\end{alignat}
Thus one can get the work density \cite{Hayward:1993wb} of the matter field in the FRW universe
\begin{alignat}{1}
W:=-\frac{1}{2}h_{ab}T^{ab}=\frac{1}{2}(\rho_m-p_m)=\frac{3}{8\pi R^2_A}-\frac{\dot{R}_A}{8\pi HR^3_A},  \label{WD}
\end{alignat}
and the thermodynamic volume is
\begin{alignat}{1}
V\equiv\frac{4\pi R_A^3}{3}. \label{TV}
\end{alignat}

With the above quantities defined, the following relation can be easily checked:
\begin{alignat}{1}
\od M=-T\od S+W\od V. \label{FL}
\end{alignat}
Compared with the first law of thermodynamics
\footnote{Strictly speaking, it is the Gibbs equation \cite{Hayward:1997jp,Wu:2009wp}.}
\begin{alignat}{1}
\od U=T\od S-P\od V, \label{SFL}
\end{alignat}
we see that the internal energy $U$ and thermodynamic pressure $P$ can be identified with $-M$ and $W$,
i.e.\
\begin{alignat}{1}
U:=&-M,
\\
P:=&W. \label{tp}
\end{alignat}
Note that this definition of $P$ using the work density is more natural than that in the literature using the cosmological constant, in the sense that it is here a true variable rather than a constant.

The equation of state for the thermodynamic pressure defined in (\ref{tp}) can be easily obtained from (\ref{KHT}) and (\ref{WD})
\begin{alignat}{1}
P=\frac{T}{2R_A}+\frac{1}{8\pi R^2_A}. \label{ES}
\end{alignat}
It is then natural to ask whether this system has a $P$-$V$ phase transition, whose necessary condition is that the equation
\begin{alignat}{1}
\left(\frac{\partial P}{\partial V}\right)_{T}=\left(\frac{\partial^2 P}{\partial V^2}\right)_{T}=0,\label{PV}
\end{alignat}
or equivalently
\begin{alignat}{1}
\left(\frac{\partial P}{\partial R_A}\right)_{T}=\left(\frac{\partial^2 P}{\partial R^2_A}\right)_{T}=0 \label{PVTc}
\end{alignat}
has a critical-point solution $T=T_c,\ P=P_c,\ R_A=R_c$.
By substituting (\ref{ES}) into (\ref{PVTc}), one can easily check that no such solution exists, and thus there is no
$P$-$V$ phase transition for the FRW universe with a perfect fluid in Einstein gravity.

\section{Modified Gravity: Equation of State}

In this section, we derive the equation of state for the FRW universe in modified gravity, and we take the gravity with a generalized conformal scalar field as an example, which belongs to the Horndeski class.

\subsection{A Brief Introduction of the Gravity with a Generalized Conformal Scalar Field}

The most generic scalar-tensor theory is Horndeski gravity \cite{Hu:2018qsy}, which allows high order derivatives in the action.
Its equations of motion has at most second order derivatives, so there are no Ostrogradsky instabilities \cite{Fernandes:2021dsb},
which is similar to Lovelock gravity. Horndeski gravity has been used to study the thermodynamics of black holes,
where $P$-$V$ phase transition has been found, and this arouses our interest that whether $P$-$V$ phase transition
can be found for the FRW universe in this gravity.

The general form of the Lagrangian in Horndeski gravity is written as \cite{Kobayashi:2019hrl}
\footnote{It is also called generalized Galileon theory \cite{Deffayet:2011gz,Kobayashi:2011nu}.},
\begin{alignat}{1}
\mathcal{L}=&G_2(\phi, X)-G_3(\phi, X)\Box\phi+G_4(\phi, X)R
+G_{4;X}[(\Box\phi)^2-\nabla^{\mu}\nabla^{\nu}\phi\nabla_{\mu}\nabla_{\nu}\phi]
+G_{5}(\phi,X)G^{\mu\nu}\nabla_{\mu}\nabla_{\nu}\phi
\nonumber \\ &
-\frac{G_{5;X}}{6}[(\Box\phi)^3-3\Box\phi\nabla^{\mu}\nabla^{\nu}\phi\nabla_{\mu}\nabla_{\nu}\phi
+2\nabla_{\mu}\nabla^{\nu}\phi\nabla_{\nu}\nabla^{\lambda}\phi\nabla_{\lambda}\nabla^{\mu}\phi], \label{gf}
\end{alignat}
where $G_2,G_3,G_4$ and $G_5$ are arbitrary functions of $\phi$ and
$X:=-\nabla^{\mu}\phi\nabla_{\mu}\phi/2\equiv-(\nabla\phi)^2/2$.
In the following, we only consider a special example of the Horndeski gravity, where the scalar field is conformally invariant \cite{Fernandes:2021dsb}.
In this case, the action is obtained as
\footnote{If $\beta$ and $\lambda$ are zero, it reduces to the action of the regularized 4d Einstein-Gauss-Bonnet gravity
\cite{Feng:2020duo}. }
\begin{alignat}{1}
S=&\frac{1}{16\pi}\left[\int(R-2\Lambda)\sqrt{-g}\od^{4}x+S_{\alpha}+S_{\beta}+S_{\lambda}+S_m\right], \label{action}
\end{alignat}
where
\begin{alignat}{1}
S_{\alpha}=&\alpha\int[2(\nabla\phi)^4+4\Box\phi(\nabla\phi)^2+4G^{\mu\nu}\nabla_{\mu}\phi\nabla_{\nu}\phi-\phi G]\sqrt{-g}\od^4x,
\\
S_{\beta}=&-\beta\int[R+6(\nabla\phi)^2]e^{2\phi}\sqrt{-g}\od^4x,
\\
S_{\lambda}=&-2\lambda\int e^{4\phi}\sqrt{-g}\od^4x,
\end{alignat}
and $S_m$ stands for the action of other matter fields, such as the perfect fluid.
The above action (\ref{action}) is a special example of Horndeski gravity (\ref{gf}) with
\begin{alignat}{1}
G_2=-2\Lambda-2\lambda e^{4\phi}+12\beta e^{2\phi}X+8\alpha X^2, \quad G_3=8\alpha X, \quad G_4=1-\beta e^{2\phi}+4\alpha X,
\quad G_5=4\alpha\log X.
\end{alignat}


\subsection{The Equation of State for the FRW Universe}

In this part, we apply this modified gravity to the FRW universe and get its equation of state.
We start with the Friedmann's equation, then get the first law of thermodynamics, and obtain the equation of state in the end.

For convenience, we apply this modified gravity (\ref{action}) to the spatially flat ($k=0$) FRW universe
with stress-tensor (\ref{ST}) and $\Lambda=0$. In this case, the modified Friedmann's equations \cite{Fernandes:2021dsb} are very simple
\footnote{Remarkably, although there are three coupling constants $\alpha,\beta,\lambda$ in the action (\ref{action}),
only $\alpha$ is present in the modified Friedmann's equations, see \cite{Fernandes:2021dsb} for the derivations.}
\begin{alignat}{1}
(1+\alpha H^2)H^2=&\frac{8\pi}{3}\rho_m, \label{fe1}
\\ (1+2\alpha H^2)\dot{H}=&-4\pi(\rho_m+p_m), \label{fe2}
\end{alignat}
and satisfy the continuity equation
\begin{alignat}{1}
\dot{\rho}_m+3H(\rho_m+p_m)=0.
\end{alignat}
Interestingly, the above equations have the same form as the ones from holographic cosmology \cite{Apostolopoulos:2008ru,Bilic:2015uol},
quantum corrected entropy-area relation \cite{Cai:2008ys}, generalized uncertainty principle \cite{Lidsey:2009xz},
and the 4d EGB gravity \cite{Feng:2020duo}.

For the spatially flat FRW universe, the relations (\ref{AH}) and (\ref{dot}) are much simplified
\begin{alignat}{1}
R_A=\frac{1}{H}, \quad \dot{R}_A=-\dot{H}R^2_A,
\end{alignat}
which can be used to rewrite the Friedmann's equations (\ref{fe1}) and (\ref{fe2}), and the results are
\begin{alignat}{1}
\left(1+\frac{\alpha}{R^2_A}\right)\frac{3}{8\pi R^2_A}=&\rho_m, \label{rhom}
\\
\left(1+\frac{2\alpha}{R^2_A}\right)\frac{\dot{R}_A}{4\pi R^2_A}=&\rho_m+p_m. \label{pm}
\end{alignat}
Therefore, the work density of the matter field is
\begin{alignat}{1}
W:=-\frac{1}{2}h_{ab}T^{ab}=\frac{1}{2}(\rho_m-p_m)=\left(1+\frac{\alpha}{R^2_A}\right)\frac{3}{8\pi R^2_A}
-\left(1+\frac{2\alpha}{R^2_A}\right)\frac{\dot{R}_A}{8\pi R^2_A}, \label{wh}
\end{alignat}
and the thermodynamic volume can still take the form of $V=4\pi R_A^3/3$.

The Kodama-Hayward temperature (\ref{KHT}) for the spatially flat FRW universe reduces to
\begin{alignat}{1}
T=\frac{1}{2\pi R_A}\left(1-\frac{\dot{R}_A}{2}\right), \label{th}
\end{alignat}
but its conjugate entropy may not take the Bekenstein-Hawking form $S=A/4$, because it is known that this form does not hold in
many theories beyond Einstein gravity (see e.g.\ \cite{Cai:2008ys}).
Corresponding to the action (\ref{action}) we make the following ansatz for the entropy
\footnote{For stationary black holes, there are a number of approaches \cite{Iyer:1995kg} to calculate the entropy,
such as the Euclidean method and Noether charge method. For dynamical black holes, Wald \cite{Wald:1993nt} proposed that
one can use locally defined geometric quantities to get the entropy. Inspired by Wald's work, Hayward \cite{Hayward:1998ee} proposed that
one can use Kodama vector instead of Killing vector in the Wald formula to give the definition of dynamical black hole entropy.
This method should be applicable to the FRW universe in this modified gravity as well, but it is more complex and will not be
used in this paper.}
\begin{alignat}{1}
S=\frac{A}{4}+\alpha f(R_A)+\beta g(R_A)+\lambda h(R_A), \label{eh}
\end{alignat}
which guarantees that the Bekenstein-Hawking entropy can be recovered in the Einstein gravity limit $\alpha,\beta,\lambda\rightarrow 0$.

The energy for the FRW universe here can be easily obtained from (\ref{rhom})
\begin{alignat}{1}
E=\rho_m V=\left(1+\frac{\alpha}{R^2_A}\right)\frac{R_A}{2},
\end{alignat}
which could be regarded as the effective Misner-Sharp energy, so it should also satisfy the relation (\ref{FL})
\begin{alignat}{1}
\od E=-T\od S+W\od V, \label{flh}
\end{alignat}
except that $W,T,S$ take the new forms in  (\ref{wh}), (\ref{th}), and (\ref{eh}) respectively.

One immediate finding is that the expression of the entropy (\ref{eh}) can be determined from (\ref{flh})
\begin{alignat}{1}
S=\frac{A}{4}+2\pi\alpha\ln\left(\frac{A}{A_0}\right), \label{entropy}
\end{alignat}
where $A_0$ can take any constant with the dimensionality of area to guarantee the logarithmic function is well defined.
In the above entropy, there is only one correction term to the Bekenstein-Hawking entropy as expected,
which is also the same form as the static black hole entropy derived in the same theory \cite{Fernandes:2021dsb},
the black hole entropy with quantum or thermal correction
\cite{Cai:2008ys,Solodukhin:1997yy,Mann:1997hm,Kaul:2000kf,Carlip:2000nv,Das:2001ic,Mukherji:2002de,
Gour:2003jj,Chatterjee:2003uv}
and the entropy of 4d Gauss-Bonnet black hole in AdS space \cite{Wei:2020poh}.

In the same way as we did for Einstein gravity, here again we identify the internal energy $U$ with $-E$
and the thermodynamic pressure $P$ with $W$, i.e.
\begin{alignat}{1}
U\equiv&-E,\\
P\equiv&W, \label{pw}
\end{alignat}
then the standard first law of thermodynamics (Gibbs equation) can be established
\begin{alignat}{1}
\od U=T\od S-P\od V.
\end{alignat}
Finally, from (\ref{wh}), (\ref{th}) and (\ref{pw}), we obtain the equation of state:
\begin{alignat}{1}
P=\frac{T}{2R_A}+\frac{1}{8\pi R^2_A}+\frac{\alpha T}{R_A^3}-\frac{\alpha}{8\pi R^4_A}.\label{esh}
\end{alignat}

\section{Modified Gravity: $P$-$V$ Phase Transition}

In this section, we show that the equation of state (\ref{esh}) for FRW universe has a critical point
and the critical exponents are the same as the mean field theory, i.e. the FRW universe has a $P$-$V$ phase transition.

For the equation of state (\ref{esh}), the critical condition (\ref{PV}) can be written as
\begin{alignat}{1}
2\pi T_c R_c^3+R_c^2+12\pi\alpha R_c T_c-2\alpha=&0,\label{1}
\\
4\pi T_c R_c^3+3R_c^2+48\pi\alpha R_c T_c-10\alpha=&0.\label{2}
\end{alignat}
If $\alpha>0$, the critical radius and temperature can not be both positive, so there is not any physical solution in this case.
If $\alpha<0$, there is a critical point
\begin{alignat}{1}
R_c=\sqrt{(6-4\sqrt{3})\alpha}, \quad T_c=\frac{\sqrt{6+4\sqrt{3}}}{12\pi\sqrt{-\alpha}}, \quad P_c=-\frac{15+8\sqrt{3}}{288\pi\alpha}.
\end{alignat}
A dimensionless constant can be acquired from the above three values:
\begin{alignat}{1}
\rho=\frac{2P_{c} R_{c}}{T_{c}}=\frac{6+\sqrt{3}}{12}.
\end{alignat}

Near the critical point, there are four critical exponents $(\tilde{\alpha},\beta,\gamma,\delta)$ defined in the following way \cite{Hu:2018qsy,Wei:2020poh}:
\begin{alignat}{1}
C_{V}=&T\left(\frac{\partial S}{\partial T}\right)_V\propto |t|^{-\tilde{\alpha}}, \label{expo}
\\
\eta=&\frac{V_l-V_s}{V_c}\propto |t|^{\beta},
\\
\kappa_T=&-\frac{1}{V}\left(\frac{\partial V}{\partial P}\right)_T\propto |t|^{-\gamma},
\\
p\propto&\quad v^{\delta},
\end{alignat}
where
\begin{alignat}{1}
t=\frac{T-T_c}{T_c}, \quad p=\frac{P-P_c}{P_c},\quad v=\frac{V-V_c}{V_c}.
\end{alignat}
In most cases, the four critical exponents satisfy the following four scaling laws
\begin{alignat}{1}
&\tilde{\alpha}+2\beta+\gamma=2,\quad \tilde{\alpha}+\beta(1+\delta)=2,
\nonumber \\
&\gamma(1+\delta)=(2-\tilde{\alpha})(\delta-1),\quad \gamma=\beta(\delta-1),\label{SL}
\end{alignat}
in which there are actually two independent relations.
In the following, we will calculate the four critical exponents and check whether they satisfy the scaling laws.

The entropy (\ref{entropy}) of the FRW universe in this modified gravity is also a function of the thermodynamic volume $V$, so $C_V$ is zero,
which means the first critical exponent $\tilde{\alpha}$ is zero. To get the other three critical exponents conveniently,
one can expand the thermodynamic pressure or the equation of state (\ref{esh}) around the critical point
\begin{alignat}{1}
p=a_{10}t+a_{11}tv+a_{03}v^3+\mathcal{O}(tv^2,v^4),\label{expansion}
\end{alignat}
where the coefficients are
\begin{alignat}{1}
a_{10}=&\left(\frac{\partial p}{\partial t}\right)_{c}=\frac{T_c}{P_c}\left(\frac{\partial P}{\partial T}\right)_{c}
=\frac{T_{c}(R_{c}^2+2\alpha)}{2P_{c} R_{c}^3}<0,
\\
a_{11}=&\left(\frac{\partial^2 p}{\partial t\partial v}\right)_{c}=\frac{R_c T_c}{3P_c}\left(\frac{\partial^2 P}{\partial T\partial R_A}\right)_c
=-\frac{T_{c}(R_{c}^2+6\alpha)}{6P_{c} R_{c}^3}>0,
\\
a_{03}=&\frac{1}{3!}\left(\frac{\partial^3 p}{\partial v^3}\right)_c=\frac{R_c^3}{162P_c}
\left(\frac{\partial^3 P}{\partial R_A^3}\right)_{c}
=\frac{R_{c}^2+6\alpha}{648\pi P_{c} R_{c}^4}<0.
\end{alignat}

The Gibbs free energy is defined as usual
\begin{alignat}{1}
G:=U+PV-TS,
\end{alignat}
so we have
\begin{alignat}{1}
\od G=-S\od T+V\od P,
\end{alignat}
and thus the Maxwell's equal area law still holds. The values of $P$ at the two endpoints
of the coexistence line are the same
\begin{alignat}{1}
p^*=a_{10}t+a_{11}tv_{s}+a_{03}v_{s}^3=a_{10}t+a_{11}tv_{l}+a_{03}v_{l}^3,
\end{alignat}
or
\begin{alignat}{1}
a_{11}(v_{l}-v_{s})t+a_{03}(v_{l}^3-v_{s}^3)=0, \label{ea1}
\end{alignat}
where the labels `s' and `l' stand for `small' and `large' respectively. Another relation is
\begin{alignat}{1}
\int v\od p=\int_{v_s}^{v_l} v\left(\frac{\partial p}{\partial v}\right)_t\od v=0,
\end{alignat}
so we have
\begin{alignat}{1}
2a_{11}(v_{l}^2-v_{s}^2)t+3a_{03}(v_{l}^4-v_{s}^4)=0. \label{ea2}
\end{alignat}
From the above two relations (\ref{ea1}) and (\ref{ea2}), one can get a nontrivial solution
\begin{alignat}{1}
v_{l}=\sqrt{-\frac{a_{11}}{a_{03}}t},\quad v_{s}=-\sqrt{-\frac{a_{11}}{a_{03}}t},
\end{alignat}
and
\begin{alignat}{1}
\eta=v_{l}-v_{s}=2\sqrt{-\frac{a_{11}}{a_{03}}t}\propto |t|^{1/2},
\end{alignat}
which shows that the second critical exponent $\beta$ is $1/2$. Interestingly, because $a_{11}>0, a_{03}<0$, we have $t>0$, which means that the coexistence phases in the $P$-$V$ diagram appear above the critical temperature $T>T_{c}$. This behavior is different from that of an AdS black hole, where coexistence phases appear below the critical temperature $T<T_{c}$.

The third critical exponent is from the isothermal compressibility near the critical point
\begin{alignat}{1}
\kappa_T=-\frac{1}{V_{c}}\left(\frac{\partial V}{\partial P}\right)_T|_{c}\propto-\left(\frac{\partial p}{\partial v}\right)^{-1}=-\frac{1}{a_{11}t}\propto t^{-1},
\end{alignat}
which shows $\gamma=1$.

The shape of the isothermal line of the critical temperature $t=0$ is
\begin{alignat}{1}
p\propto v^3,
\end{alignat}
which provides the fourth critical exponent $\delta=3$.

In summary, the four critical exponents are:
\begin{alignat}{1}
\tilde{\alpha}=0,\quad \beta=\frac{1}{2},\quad \gamma=1, \quad \delta=3,
\end{alignat}
which are the same as those in the mean field theory and satisfy the scaling laws (\ref{SL}).

\section{Conclusions and Discussions}

In this paper, we have studied the thermodynamic properties, especially the equation of state and $P$-$V$ phase transitions
of the FRW universe with a perfect fluid in Einstein gravity and a modified theory of gravity that belongs to the Horndeski class.
The thermodynamic pressure $P$ of the FRW universe is defined as the work density $W$, which is a natural definition directly read out from the first law of thermodynamics. We have derived the equations of state, and impressively in the modified gravity case, it exhibits $P$-$V$ phase transitions. To our best knowledge of the literature, this is the first time that such phase transitions are found in a spacetime that is not asymptotically AdS black holes. The phase transitions occur above the critical temperature, which is different from the AdS black holes. In the end, we have calculated the four critical exponents, which are the same as those in the mean field theory and thus satisfy the scaling laws.

We would like to discuss a few more open questions related to our work. The first natural question is whether $P$-$V$ phase transitions can be found for FRW universe in other modified theories of gravity and/or filled with other fields.
The second question is whether $P$-$V$ phase transitions can be found in black holes inside the FRW universe
\footnote{For the McVittie black hole, a Hawking-Page-like phase transition instead of a $P$-$V$ phase transition
has been found in Einstein gravity \cite{Abdusattar:2022bpg}.} and other dynamical black holes.
We will carry these investigations in the future.

\section{Acknowledgment}

We are grateful for the valuable comments from the referees and useful discussions with Li-Ming Cao,
Shao-Wen Wei, Yen Chin Ong, Wen-Long You, et al. This work is supported by the National Natural Science
Foundation of China (NSFC) under grants Nos. 12175105, 11575083, 11565017, Top-notch Academic Programs Project of Jiangsu Higher
Education Institutions (TAPP), and “the Fundamental Research Funds for the Central Universities, No. NS2020054” of China. H. Zhang is
supported by the National Natural Science Foundation of China Grant No.12235019, and the National Key Research and Development Program
of China (No. 2020YFC2201400).

\appendix

\section{The Quasi-equilibrium Properties of the FRW Universe}

One can use the physical radius $R:=a(t)r$  instead of the comoving coordinate $r$
and get a new line element for the FRW universe \cite{Cai:2008gw}
\begin{alignat}{1}
\od s^2=-\frac{1-R^2/R_A^2}{1-k R^2/a^2}\od t^2-\frac{2HR}{1-kR^2/a^2}\od t\od R
+\frac{1}{1-kR^2/a^2}\od R^2+R^2(\od\theta^2+\sin^2\theta\od\varphi^2), \label{nle}
\end{alignat}
which is quite similar to the Painleve-de Sitter metric \cite{Parikh:2002qh} in the case $k=0$.
If one demands the Hubble parameter $H$ to be a constant as well as $k=0$, it is just the
Painleve-de Sitter metric. If $H$ changes very slowly, the FRW universe is
a quasi-de Sitter space.

For the FRW universe as well as other dynamical spacetimes,
there exists a counterpart of the Killing vector field, which is the Kodama vector field.
For spacetimes with spherical symmetry, the Kodama vector is defined as
\cite{Kodama:1979vn,Minamitsuji:2003at,Maeda:2007uu,Cai:2008gw,Cai:2009qf}
\begin{equation}
K^{a}:=-\epsilon^{ab}\nabla_{b} R.
\end{equation}
For the above line element (\ref{nle}), the Kodama vector can be obtained as
\begin{alignat}{1}
K^a=\sqrt{1-k\frac{R^2}{a^2}}\left(\frac{\partial}{\partial t}\right)^a,
\end{alignat}
which is very similar to the Killing vector $(\partial/\partial t)^a$ in de Sitter space.

With the Kodama vector field, one can define conserved quantities such as the Misner-Sharp energy
for the FRW universe. At first, one can define the energy current
\begin{equation}
J^{\mu}:=-T^{\mu}_{~~\nu}K^{\nu},
\end{equation}
which is divergence free
\begin{equation}
\nabla_{\mu}J^{\mu}=0
\end{equation}
in Einstein gravity, and many modified theories of gravity.
Then, one can define the associated conserved charge (generalized Misner-Sharp energy)
\begin{equation}
Q_J:=\frac{1}{8\pi}\int_{\Sigma} J^{\mu}\mathrm{d}\Sigma_{\mu},
\end{equation}
where $\Sigma$ is a hypersurface and $\mathrm{d}\Sigma_{\mu}$ is its directed surface element.
The Misner-Sharp energy satisfies the unified first law \cite{Hayward:1997jp,Cai:2006rs}, which after being projected onto the apparent horizon leads to the first law of thermodynamics. One can find that, for the FRW universe in Einstein gravity and some modified theories of gravity, the Clausius relation holds,
which also suggests that the FRW universe is an (quasi-)equilibrium thermodynamic system \cite{Cai:2006pa,Wu:2009wp}.

\end{document}